\documentclass[useAMS,usenatbib]{mn2e}
\usepackage{graphicx}
\usepackage{amssymb}
\usepackage[T1]{fontenc}
\usepackage{aecompl}

\newcommand{\mnras}{{MNRAS}}
\newcommand{\apj}{{ApJ}}
\newcommand{\aap}{{A\&A}}

\newcommand{\apjl}{{ApJL}}
\newcommand{\apjs}{{ApJS}}
\newcommand{\aj}{{AJ}}

\newcommand{\ie}{{i.e.}\,}
\newcommand{\eg}{{e.g.}\,}
\renewcommand{\fdg}{\mbox{\ensuremath{.\!\!^\circ}}}
\renewcommand{\farcs}{\mbox{\ensuremath{.\!\!^{\prime\prime}}}}

\title[The CASSOWARY lensing systems]{The evolution of late-type galaxies from CASSOWARY lensing systems}

\author[Zuzanna Kostrzewa-Rutkowska et~al.]{Zuzanna Kostrzewa-Rutkowska$^{1}$\thanks{E-mail: zkostrzewa@astrouw.edu.pl}, {\L}ukasz Wyrzykowski$^{1,2}$, Matthew W.\ Auger$^{2}$,  \newauthor Thomas E.\ Collett$^{2}$, Vasily Belokurov$^{2}$\\
$^{1}$Warsaw University Astronomical Observatory, Al.\ Ujazdowskie 4, 00-478 Warszawa, Poland\\
$^{2}$Institute of Astronomy, University of Cambridge, Madingley Road, Cambridge CB3 0HA, UK} 

\begin{document}

\date{2014 Apr 10}

\pagerange{\pageref{firstpage}--\pageref{lastpage}} \pubyear{2014}

\maketitle

\label{firstpage}

\begin{abstract}
We explore the properties of lensing galaxies and lensed faint sources at redshifts between 1.5 and 3.0.
Our sample consists of 9 carefully selected strongly-lensed galaxies discovered by the CASSOWARY survey in the Sloan Digital Sky Survey (SDSS) data.
We find that, despite some limitations of the original SDSS data, the homogeneous sample of lensing systems can provide a useful insight into lens and source properties.
We also explore the limitations of using low-resolution data to model and analyse galaxy-galaxy lensing.
We derive the relative alignment of mass and light in fitted lens profiles.
The range of magnification extends above 5, hence we are able to analyse potentially small and low-mass galaxies at high redshifts.
We confirm the likely evolution of the size-luminosity relation for blue star-forming galaxies as a function of redshift.
\end{abstract}


\section{Introduction}
Gravitational lensing is a powerful tool; acting as a natural telescope it can potentially allow astronomical observations to reach smaller and fainter objects than would otherwise be observable (\citealt*{1992grle.book.....S}; \citealt{2006glsw.conf....1S}).  
In addition to producing multiple images the source is significantly magnified; typically by a factor of $\sim$10 (\eg \citealt{2011ApJ...734..104N}).
Strong lenses have been used, for example, to investigate the density profile of lensing galaxies (\eg \citealt{2008ApJ...682..964B}; \citealt{2010ApJ...721L.163A}; \citealt{2011ApJ...727...96R}), the evolution of galaxies (\eg \citealt{2011ApJ...734..104N}; \citealt{2012ApJ...757...82B}) and cosmological parameters (\eg \citealt{2012ApJ...750...10S}; \citealt{2012MNRAS.424.2864C}; \citealt{2013ApJ...766...70S}).

The redshift range $z \sim 1-3$ is a crucial period in the evolution of galaxies when today-observed galaxies formed the majority of their stars.
Studies have shown that star formation generally occurred in galaxies with irregular and multicomponent morphologies (\eg \citealp{2012ApJ...745...85L}).
Modern models of galaxy evolution produce relations between the fundamental physical properties of galaxies (\ie size, mass, luminosity).
Our homogeneous sample with similar photometric properties and redshifts allows us to explore these relations for blue, star-forming, high-redshift galaxies.
At the same time, we are able to explore the properties of deflectors - early type galaxies at redshifts $z \sim 0.2-0.7$ probing the relative alignment of mass and light in flattening and orientation.
In this paper we explore in detail lens and source properties of lensing systems selected from the CASSOWARY catalogue for which we obtained reliable models. 
We also study and show the advantages and limitations of the available imaging data. 

This paper is organized as follows. 
In Section 2 we introduce our CASSOWARY sample and present selection method, then describe the lens modelling in Section 3. 
In Section 4 we discuss results of the modelling and lens and source properties. 
We discuss the results and conclude in Section 5.

Throughout this paper we assume a flat $\Lambda$-Cold-Dark-Matter ($\Lambda$CDM) cosmological model of the Universe with parameters $\Omega_\Lambda=0.73$, $\Omega_\mathrm{M}=0.27$ and $H_0=71 \mathrm{km~s^{-1} Mpc^{-1}}$, $h=0.71$ as favoured by the seven year Wilkinson Microwave Anisotropy Probe (WMAP) results (\citealt{2011ApJS..192...18K}).

\section{Observations \& system selection}
\label{sec:data}

\begin{table*}
\centering
\caption{The lensing systems selected from the CASSOWARY catalogue for this study. All redshifts $z_L$ and $z_S$ for foreground and background galaxies respectively are taken from \citet{2013MNRAS.436.1040S} except CSWA 161 (for this system we use the photometric redshift from the SDSS data base).}
\begin{tabular}{l c l l l l}
\hline
SDSS ID & CSWA ID & RA (deg) & Dec (deg) & $z_L$ & $z_S$ \\
\hline
SDSS J1148+1930 & 1 & 177.1380745 & 19.50087261 & 0.444 & 2.379 \\
SDSS J1206+5142 & 6 & 181.50871224 & 51.70820438 & 0.433 & 2.000 \\
SDSS J0022+1431 & 21 & 5.67049154 & 14.51956525 & 0.380 & 2.730 \\
SDSS J0145--0455 & 103 & 26.26788977 & -4.93099739 & 0.633 & 1.958 \\
SDSS J0143+1607 & 116 & 25.95886849 & 16.12749833 & 0.415 & 1.499 \\
SDSS J2340+2947 & 161 & 355.11895889 & 29.79641554 & 0.497 & -- \\
SDSS J2158+0257 & 163 & 329.68199579 & 2.95839038 & 0.285 & 2.081 \\
SDSS J0232--0323 & 164 & 38.20778268 & -3.39056271 & 0.450 & 2.518 \\
SDSS J0105+0144 & 165 & 16.33188373 & 1.7489944 & 0.361 & 2.127 \\
\hline
\end{tabular}
\label{tab:CSWAsample}
\end{table*}

The Cambridge And Sloan Survey Of Wide ARcs in the skY (CASSOWARY)\footnote{http://www.ast.cam.ac.uk/ioa/research/cassowary/} discovered lenses by searching for multiple blue companions (or arcs) separated by $\gtrsim 3$ arcseconds from luminous red galaxies (\citealp{2009MNRAS.392..104B}).
Typically the CASSOWARY lenses are early type galaxies or groups of galaxies at $z \sim 0.2-0.7$ and the CASSOWARY sources are $z \sim 1-3$ star-forming galaxies. 
The current CASSOWARY catalogue includes more than 50 spectroscopically confirmed strong lenses. 
Thanks to their large image separations the systems are well resolved with SDSS imaging (\citealp{2011ApJS..193...29A}). 
The CASSOWARY lenses make an ideal homogeneous sample of systems for studying the statistical properties of the source galaxies with original survey data without follow-up.

The selection procedure consists of two parts. 
First, a broad search for all massive ellipticals with at least one blue companion was carried out.
Secondly, a number of more targeted searches for arcs of different sizes and brightnesses on the initial catalogue of candidates was performed.
To obtain the initial sample, a variant of the photometric criteria was used to identify luminous red galaxies in SDSS's Galaxy table.
The sample only includes galaxies with $g - r < 2.5$ and $r - i < 1.5$. 
The arcs are required to be bright (on average brighter than 21.5 mag) and to have at least two components of comparable magnitude. 

Although wide-separation lenses are comparatively unexplored, they are interesting for a number of reasons. 
First, wide-separation lenses probe the high-mass end of the galaxy mass function since the Einstein radius increases with lens mass. Since the Einstein radius increases more rapidly with mass than the half light radius of the galaxy, the Einstein radius of wide-separation lenses is typically a few effective radii where the matter distribution is dominated by dark matter.
Secondly since the source is often highly magnified, the CASSOWARY sample provides us with a sample of the brightest observed high redshifts galaxies. Because of the high magnification and the fact that the images form well outside the effective radius of the lensing galaxy source properties, particularly observed source brightness, can be measured with high accuracy. 

From the CASSOWARY sample, we selected systems that are characterized by large separations, one dominant lensing galaxy and the absence of blending stars.
In this paper we made use of the SDSS data in four bands ($griz$) to perform the modelling of the lens systems.
In Table \ref{tab:CSWAsample} we list the 9 selected systems from the CASSOWARY catalogue.

\section{Mass and surface brightness modelling}
We model the lens surface mass distribution as a singular isothermal ellipsoid (SIE; \eg \citealp{1994A&A...284..285K}) and include external shear.
Several authors have shown that the SIE model provides a good approximation of the lens potential on galaxy scales (\citealp{2004ApJ...611..739T}; \citealp{2005ApJ...623..666R}; \citealp{2006ApJ...649..599K}).
We use the method described in \citet{2011MNRAS.411L...6A} and extend it to fit multiple images---in multiple filters---simultaneously (\ie \citealp{2013MNRAS.436..503A}). 
This extension significantly improves the model constraints because CASSOWARY lenses are typically red whilst CASSOWARY sources tend to be blue. 

\begin{table*}
\centering
\caption{Model parameters with assumed limits and prior. (Note: $1 \mathrm{pix}=0\farcs396$).}
\begin{tabular}{l l l l}
\hline 
 & Parameter & Prior & Limits \\
\hline
Lens mass profile & $x_M, y_M$ (pix)  & Uniform & [-5;5] \\
 & $\theta_E$ (pix) & Uniform & [2;180] \\
 & $q_M$ & Uniform & [0.2;1.0] \\
 & $\theta_M$ (rad) & Uniform & [0;2$\pi$] \\
 & $\gamma$ & Uniform & [0.0;0.5] \\ 
 & $\theta_{\gamma}$ (rad) & Uniform & [$-\pi$;$\pi$] \\
\hline
Lens light profile & $x_*, y_*$ (pix) & Normal & ($\mu=x_M, \sigma=3$), ($\mu=y_M, \sigma=3$) \\
 & $R_*$ (pix) & Uniform & [1;100] \\
 & $q_*$ & Truncated Normal & [0.0;1.0], $\mu=q_M, \sigma=0.15$ \\
 & $\theta_*$ (rad) & Truncated Normal & [0;2$\pi$], $\mu=\theta_M, \sigma=20$ \\
 & $n_L$ & Uniform & [0.5;8.0] \\
\hline
Source light profile & $x_S, y_S$ (pix) & Uniform & [$x_M-20$;$x_M+20$], [$y_M-20$;$y_M+20$] \\
 & $R_S$ (pix) & Uniform & [1;100] \\
 & $q_S$ & Uniform & [0.2;1.0] \\
 & $\theta_S$ (rad) & Uniform & [$-\pi$;$\pi$] \\
 & $n_S$ & Uniform & [0.5;8.0] \\
\hline
Offset & $x_{gr}$, $y_{gr}$ (pix) & Uniform & [-3;3] \\
 & $x_{gi}$, $y_{gi}$ (pix) & & \\
 & $x_{gz}$, $y_{gz}$ (pix) & & \\
\hline
Noise parameter & $A_{griz}$ & Uniform & [1;10] \\ 
\hline
\end{tabular}
\label{tab:param}
\end{table*}

The lens model has seven parameters (five for the SIE and two for the external shear): the Einstein radius $\theta_E$, the elliptical axis ratio $q_M=b/a$ and position angle $\theta_M$, and the position of the lens $(x,y)$; the value of shear $\gamma$ and the positional angle of the shear $\theta_\gamma$. 
The light of the lens and source galaxies were modelled with an elliptical S\'ersic profile with S\'ersic index $n_L$ and $n_S$ free to vary between 0.5 and 8. 
We do not assume that the lens mass and lens light are strongly correlated; we do not assume the mass and light are concentric, nor that they have the same ellipticity.
We chose to model the source galaxy with a simply parametrized S\'ersic profile because we are interested in average properties of galaxies (luminosity, mass, size) as a function of redshift.
Moreover, because the SDSS imaging is not high resolution the source images have typical sizes of only a few pixels, thus there is not sufficient information to justify more complicated models.

We use the imaging in four bands (SDSS $griz$) to model the systems. 
The lens and source parameters are fixed between filters.
While the lens and source colours are free. 
The parameters of our model are listed in Table \ref{tab:param}, along with the prior probability distribution functions (PDF) assigned to them and the range over which they are allowed to vary.
We convolve the models with a synthetic PSF and compute the residuals between the convolved models and the data.

To find the best fitting models for each system we use the same optimization method as in \citet{2011MNRAS.411L...6A}.
We then estimate the uncertainties of our models using the MCMC ensemble sampler {\sc emcee}\footnote{http://dan.iel.fm/emcee/}. 
However, these uncertainties are only meaningful under the assumption that the model perfectly describes the underlying system; \ie, the model insists that the source light profile is S\'ersic and the lens is SIE. 
Real galaxies are more complicated so we up-weight the full noise maps by a multiplicative constant (one per filter), which we include as a non-linear parameter in our MCMC sampling. 
Allowing the data to upweight the noise lets the data inform us how over simplified the model is\footnote{For example, if the model is very far from reality, the code should upweight to noise to the point where the recovered posteriors are the same as the initial priors. If the model is sufficiently close to reality the multiplicative constant will be close to 1.}.
From the {\sc emcee} samples we report the median and 68 per cent confidence interval of each parameter. 
We also use the samples to compute photometric magnitudes for the galaxy and source and the magnification of the source.

\section{Results}

\begin{table*}
\caption{Modelling parameters for 9 CSWA systems. (Note: The parameters $\theta_M$, $\theta_{\gamma}$, $\theta_*$ and $\theta_S$ give major-axis position  angles measured east from north.)}
\centering
\resizebox{\textwidth}{!}{
\begin{tabular}{l c c c c c c c c c}
\hline
 & \multicolumn{3}{c}{mass} & \multicolumn{3}{|c}{lens} & \multicolumn{3}{|c}{source} \\
\hline
CSWA ID & $\theta_E~['']$ & $q_M$ $\theta_M$ [deg] & $\gamma$ $\theta_{\gamma}$ [deg] & $R_*~['']$ & $q_*$ $\theta_*$ [deg] & $n_L$ & $R_S~['']$ & $q_S$ $\theta_S$ [deg] & $n_S$ \\
\hline
CSWA 1 & $ 5.08 ^{+0.02 }_{-0.02 }$& $ 0.89 ^{+0.03 }_{-0.04 }$ $ 130 ^{+8 }_{-14 }$& $ 0.021 ^{+0.012 }_{-0.012 }$ $ -14 ^{+19 }_{-13 }$& $ 6.62 ^{+1.60 }_{-1.58 }$& $ 0.90 ^{+0.04 }_{-0.03 }$ $ 133 ^{+10 }_{-10 }$& $ 7.17 ^{+0.56 }_{-0.79 }$& $ 0.81 ^{+0.48 }_{-0.33 }$& $ 0.82 ^{+0.11 }_{-0.14 }$ $ -26 ^{+35 }_{-22 }$& $ 6.85 ^{+0.85 }_{-2.17 }$ \\
CSWA 6 & $ 3.86 ^{+0.06 }_{-0.04 }$& $ 0.68 ^{+0.12 }_{-0.14 }$ $ 103 ^{+3 }_{-3 }$& $ 0.037 ^{+0.017 }_{-0.019 }$ $ 152 ^{+21 }_{-26 }$& $ 3.98 ^{+0.59 }_{-0.50 }$& $ 0.83 ^{+0.02 }_{-0.03 }$ $ 106 ^{+5 }_{-5 }$& $ 3.87 ^{+0.34 }_{-0.34 }$& $ 0.21 ^{+0.11 }_{-0.07 }$& $ 0.57 ^{+0.17 }_{-0.08 }$ $ -7 ^{+5 }_{-8 }$& $ 1.13 ^{+0.25 }_{-0.58 }$ \\
CSWA 21 & $ 3.52 ^{+0.06 }_{-0.06 }$& $ 0.48 ^{+0.07 }_{-0.04 }$ $ 12 ^{+1 }_{-1 }$& $ 0.017 ^{+0.024 }_{-0.013 }$ $ 23 ^{+64 }_{-15 }$& $ 10.38 ^{+2.08 }_{-2.47 }$& $ 0.71 ^{+0.02 }_{-0.03 }$ $ 12 ^{+3 }_{-3 }$& $ 7.27 ^{+0.50 }_{-0.72 }$& $ 0.19 ^{+0.02 }_{-0.03 }$& $ 0.52 ^{+0.10 }_{-0.11 }$ $ 82 ^{+6 }_{-7 }$& $ 1.12 ^{+0.65 }_{-0.34 }$ \\
CSWA 103 & $ 1.87 ^{+0.14 }_{-0.12 }$& $ 0.61 ^{+0.20 }_{-0.18 }$ $ 34 ^{+14 }_{-15 }$& $ 0.214 ^{+0.063 }_{-0.081 }$ $ 51 ^{+9 }_{-7 }$& $ 5.09 ^{+3.50 }_{-2.17 }$& $ 0.60 ^{+0.07 }_{-0.06 }$ $ 43 ^{+5 }_{-6 }$& $ 5.60 ^{+1.52 }_{-1.47 }$& $ 0.65 ^{+0.16 }_{-0.09 }$& $ 0.48 ^{+0.11 }_{-0.10 }$ $ -159 ^{+9 }_{-11 }$& $ 1.18 ^{+0.68 }_{-0.40 }$ \\
CSWA 116 & $ 2.62 ^{+0.03 }_{-0.03 }$& $ 0.87 ^{+0.08 }_{-0.08 }$ $ 99 ^{+22 }_{-19 }$& $ 0.110 ^{+0.031 }_{-0.024 }$ $ 169 ^{+6 }_{-6 }$& $ 12.66 ^{+4.81 }_{-3.94 }$& $ 0.84 ^{+0.06 }_{-0.05 }$ $ 147 ^{+9 }_{-11 }$& $ 6.71 ^{+0.81 }_{-0.95 }$& $ 0.56 ^{+0.30 }_{-0.14 }$& $ 0.89 ^{+0.08 }_{-0.15 }$ $ 7 ^{+53 }_{-46 }$& $ 2.51 ^{+1.36 }_{-0.98 }$ \\
CSWA 161 (L1) & $ 2.67 ^{+0.13 }_{-0.12 }$& $ 0.31 ^{+0.05 }_{-0.04 }$ $ 144 ^{+2 }_{-2 }$& $ 0.027 ^{+0.020 }_{-0.017 }$ $ -5 ^{+7 }_{-7 }$& $ 5.44 ^{+1.20 }_{-0.92 }$& $ 0.53 ^{+0.05 }_{-0.05 }$ $ 143 ^{+4 }_{-3 }$& $ 7.34 ^{+0.48 }_{-0.63 }$& $ 0.19 ^{+0.12 }_{-0.07 }$& $ 0.40 ^{+0.17 }_{-0.12 }$ $ 60 ^{+9 }_{-10 }$& $ 6.33 ^{+1.19 }_{-1.37 }$ \\
CSWA 161 (L2) & $ 3.31 ^{+0.27 }_{-0.26 }$& $ 0.81 ^{+0.11 }_{-0.12 }$ $ 133 ^{+13 }_{-14 }$ \\
CSWA 163 & $ 3.36 ^{+0.04 }_{-0.05 }$& $ 0.88 ^{+0.08 }_{-0.08 }$ $ 125 ^{+23 }_{-15 }$& $ 0.146 ^{+0.018 }_{-0.015 }$ $ -96 ^{+5 }_{-5 }$& $ 20.12 ^{+1.20 }_{-1.39 }$& $ 0.89 ^{+0.02 }_{-0.02 }$ $ 91 ^{+6 }_{-6 }$& $ 7.86 ^{+0.11 }_{-0.18 }$& $ 0.17 ^{+0.12 }_{-0.06 }$& $ 0.65 ^{+0.17 }_{-0.21 }$ $ -26 ^{+27 }_{-18 }$& $ 5.57 ^{+1.72 }_{-3.31 }$ \\
CSWA 164 & $ 3.68 ^{+0.02 }_{-0.02 }$& $ 0.88 ^{+0.05 }_{-0.09 }$ $ 164 ^{+6 }_{-5 }$& $ 0.019 ^{+0.010 }_{-0.008 }$ $ -44 ^{+13 }_{-29 }$& $ 7.62 ^{+2.40 }_{-1.95 }$& $ 0.88 ^{+0.05 }_{-0.04 }$ $ 177 ^{+10 }_{-58 }$& $ 6.65 ^{+0.77 }_{-0.87 }$& $ 0.16 ^{+0.07 }_{-0.06 }$& $ 0.42 ^{+0.08 }_{-0.12 }$ $ -44 ^{+16 }_{-8 }$& $ 0.90 ^{+5.34 }_{-0.34 }$\\
CSWA 165 & $ 3.77 ^{+0.11 }_{-0.16 }$& $ 0.64 ^{+0.11 }_{-0.11 }$ $ 83 ^{+13 }_{-18 }$& $ 0.092 ^{+0.068 }_{-0.065 }$ $ -63 ^{+18 }_{-11 }$& $ 5.04 ^{+0.89 }_{-0.77 }$& $ 0.74 ^{+0.02 }_{-0.02 }$ $ 113 ^{+2 }_{-3 }$& $ 5.15 ^{+0.42 }_{-0.41 }$& $ 0.80 ^{+1.02 }_{-0.24 }$& $ 0.68 ^{+0.23 }_{-0.26 }$ $ 37 ^{+32 }_{-33 }$& $ 2.19 ^{+2.71 }_{-1.06 }$ \\
\hline
\end{tabular}
}
\label{tab:param1}
\end{table*}

\begin{table*}
\caption{Lens and source photometry in $griz$ bands, total magnification in lens systems and source stellar mass.}
\centering
\resizebox{\textwidth}{!}{
\begin{tabular}{l c c c c}
\hline
CSWA ID & lens photometry ($griz$) & source photometry ($griz$) & magnification & $\textrm{M}_* [\textrm{M}_\odot]$\\
\hline
CSWA 1 & $ 20.16 ^{+0.14 }_{-0.12 }$ $ 18.29 ^{+0.13 }_{-0.12 }$ $ 17.51 ^{+0.13 }_{-0.12 }$ $ 17.05 ^{+0.14 }_{-0.12 }$& $ 23.25 ^{+0.27 }_{-0.30 }$ $ 23.07 ^{+0.28 }_{-0.30 }$ $ 23.06 ^{+0.29 }_{-0.30 }$ $ 22.72 ^{+0.27 }_{-0.29 }$& $ 21.11 ^{+4.96 }_{-4.35 }$ & $10.687 \pm 0.108$\\ 
CSWA 6 & $ 19.91 ^{+0.08 }_{-0.09 }$ $ 18.21 ^{+0.08 }_{-0.09 }$ $ 17.51 ^{+0.08 }_{-0.08 }$ $ 17.10 ^{+0.08 }_{-0.08 }$& $ 23.27 ^{+0.35 }_{-0.44 }$ $ 22.91 ^{+0.36 }_{-0.45 }$ $ 22.73 ^{+0.37 }_{-0.46 }$ $ 22.48 ^{+0.38 }_{-0.50 }$& $ 17.54 ^{+6.33 }_{-5.84 }$ & $10.830 \pm 0.282$\\ 
CSWA 21 & $ 19.44 ^{+0.14 }_{-0.10 }$ $ 17.75 ^{+0.14 }_{-0.10 }$ $ 17.06 ^{+0.14 }_{-0.10 }$ $ 16.61 ^{+0.14 }_{-0.10 }$& $ 22.87 ^{+0.13 }_{-0.10 }$ $ 22.47 ^{+0.13 }_{-0.10 }$ $ 22.28 ^{+0.14 }_{-0.10 }$ $ 22.31 ^{+0.14 }_{-0.10 }$& $ 12.49 ^{+1.49 }_{-1.00 }$ & $10.790 \pm 0.132$\\ 
CSWA 103 & $ 21.67 ^{+0.37 }_{-0.36 }$ $ 19.81 ^{+0.31 }_{-0.29 }$ $ 18.65 ^{+0.31 }_{-0.29 }$ $ 18.03 ^{+0.32 }_{-0.29 }$& $ 22.68 ^{+0.17 }_{-0.23 }$ $ 22.35 ^{+0.18 }_{-0.23 }$ $ 22.40 ^{+0.20 }_{-0.25 }$ $ 23.18 ^{+0.46 }_{-0.39 }$& $ 5.83 ^{+0.82 }_{-0.80 }$ & $10.661 \pm 0.356$\\ 
CSWA 116 & $ 20.04 ^{+0.22 }_{-0.21 }$ $ 18.20 ^{+0.21 }_{-0.18 }$ $ 17.42 ^{+0.21 }_{-0.18 }$ $ 16.96 ^{+0.21 }_{-0.18 }$& $ 22.63 ^{+0.25 }_{-0.27 }$ $ 22.46 ^{+0.27 }_{-0.27 }$ $ 22.18 ^{+0.28 }_{-0.28 }$ $ 22.07 ^{+0.27 }_{-0.27 }$& $ 10.28 ^{+2.10 }_{-1.79 }$ & $10.263 \pm 0.125$\\ 
CSWA 161 & $ 21.32 ^{+0.13 }_{-0.11 }$ $ 19.20 ^{+0.10 }_{-0.10 }$ $ 18.22 ^{+0.11 }_{-0.10 }$ $ 17.75 ^{+0.10 }_{-0.10 }$& $ 23.11 ^{+0.19 }_{-0.20 }$ $ 23.02 ^{+0.20 }_{-0.21 }$ $ 22.80 ^{+0.21 }_{-0.19 }$ $ 22.66 ^{+0.26 }_{-0.23 }$& $ 6.98 ^{+0.98 }_{-0.90 }$& ---\\ 
CSWA 163 & $ 18.47 ^{+0.06 }_{-0.06 }$ $ 16.82 ^{+0.05 }_{-0.05 }$ $ 16.20 ^{+0.05 }_{-0.05 }$ $ 15.84 ^{+0.05 }_{-0.05 }$& $ 23.49 ^{+0.24 }_{-0.30 }$ $ 23.40 ^{+0.25 }_{-0.31 }$ $ 23.35 ^{+0.26 }_{-0.30 }$ $ 24.80 ^{+0.39 }_{-0.47 }$& $ 14.22 ^{+2.62 }_{-2.52 }$ & $9.512 \pm 0.294$\\ 
CSWA 164 & $ 19.66 ^{+0.16 }_{-0.14 }$ $ 18.01 ^{+0.16 }_{-0.14 }$ $ 17.23 ^{+0.16 }_{-0.13 }$ $ 16.82 ^{+0.16 }_{-0.13 }$& $ 23.63 ^{+0.32 }_{-0.31 }$ $ 23.52 ^{+0.32 }_{-0.31 }$ $ 23.55 ^{+0.33 }_{-0.32 }$ $ 23.67 ^{+0.34 }_{-0.32 }$& $ 34.65 ^{+11.16 }_{-7.61 }$& $9.553 \pm 0.195$\\
CSWA 165 & $ 19.82 ^{+0.10 }_{-0.10 }$ $ 17.95 ^{+0.10 }_{-0.09 }$ $ 17.33 ^{+0.10 }_{-0.09 }$ $ 16.92 ^{+0.10 }_{-0.09 }$& $ 22.82 ^{+0.34 }_{-0.40 }$ $ 22.54 ^{+0.36 }_{-0.40 }$ $ 22.24 ^{+0.39 }_{-0.40 }$ $ 22.50 ^{+0.46 }_{-0.46 }$& $ 5.99 ^{+1.71 }_{-1.38 }$& $10.129 \pm 0.296$\\
\hline
\end{tabular}
}
\label{tab:param2}
\end{table*}

\begin{table*}
\caption{Noise parameters $A_{griz}$ with uncertainties.}
\begin{tabular}{l c c c c}
\hline
CSWA ID & $A_g$ & $A_r$ & $A_i$ & $A_z$ \\
\hline
CSWA 1 & 1.18 $\pm$ 0.02 & 1.21 $\pm$ 0.02 & 1.30 $\pm$ 0.03 & 1.22 $\pm$ 0.02 \\
CSWA 6 & 1.25 $\pm$ 0.03 & 1.21 $\pm$ 0.02 & 1.35 $\pm$ 0.03 & 1.26 $\pm$ 0.02 \\
CSWA 21 & 1.18 $\pm$ 0.02 & 1.22 $\pm$ 0.02 & 1.16 $\pm$ 0.02 & 1.21 $\pm$ 0.02 \\
CSWA 103 & 1.18 $\pm$ 0.02 & 1.19 $\pm$ 0.02 & 1.12 $\pm$ 0.02 & 1.23 $\pm$ 0.02 \\
CSWA 116 & 1.20 $\pm$ 0.02 & 1.27 $\pm$ 0.02 & 1.18 $\pm$ 0.02 & 1.19 $\pm$ 0.02 \\
CSWA 161 & 1.18 $\pm$ 0.02 & 1.28 $\pm$ 0.02 & 1.32 $\pm$ 0.02 & 1.30 $\pm$ 0.02 \\
CSWA 163 & 1.23 $\pm$ 0.02 & 1.46 $\pm$ 0.03 & 1.49 $\pm$ 0.03 & 1.26 $\pm$ 0.03 \\
CSWA 164 & 1.26 $\pm$ 0.02 & 1.22 $\pm$ 0.02 & 1.28 $\pm$ 0.02 & 1.19 $\pm$ 0.03 \\
CSWA 165 & 1.23 $\pm$ 0.02 & 1.18 $\pm$ 0.02 & 1.16 $\pm$ 0.02 & 1.23 $\pm$ 0.02 \\
\hline
\end{tabular}
\label{tab:AAA}
\end{table*}

We make an attempt to model 9 CASSOWARY systems with singular isothermal ellipsoid lens models fitted to the SDSS data.
Three of them were previously described in the literature and we used them to estimate the goodness of our fits: CSWA 1, 6, 21 (note that not all parameters are available).
After the comparison of already published and re-modelled systems we also show 6 modelled systems, not yet presented in the literature.
For the whole sample of 9 systems we present lens and source properties.
In these cases the quality of the SDSS data did not allow us to model the lens systems robustly.
The rejected systems were those where we could not robustly identify a counter image.

\begin{figure*}
\includegraphics[scale=0.55]{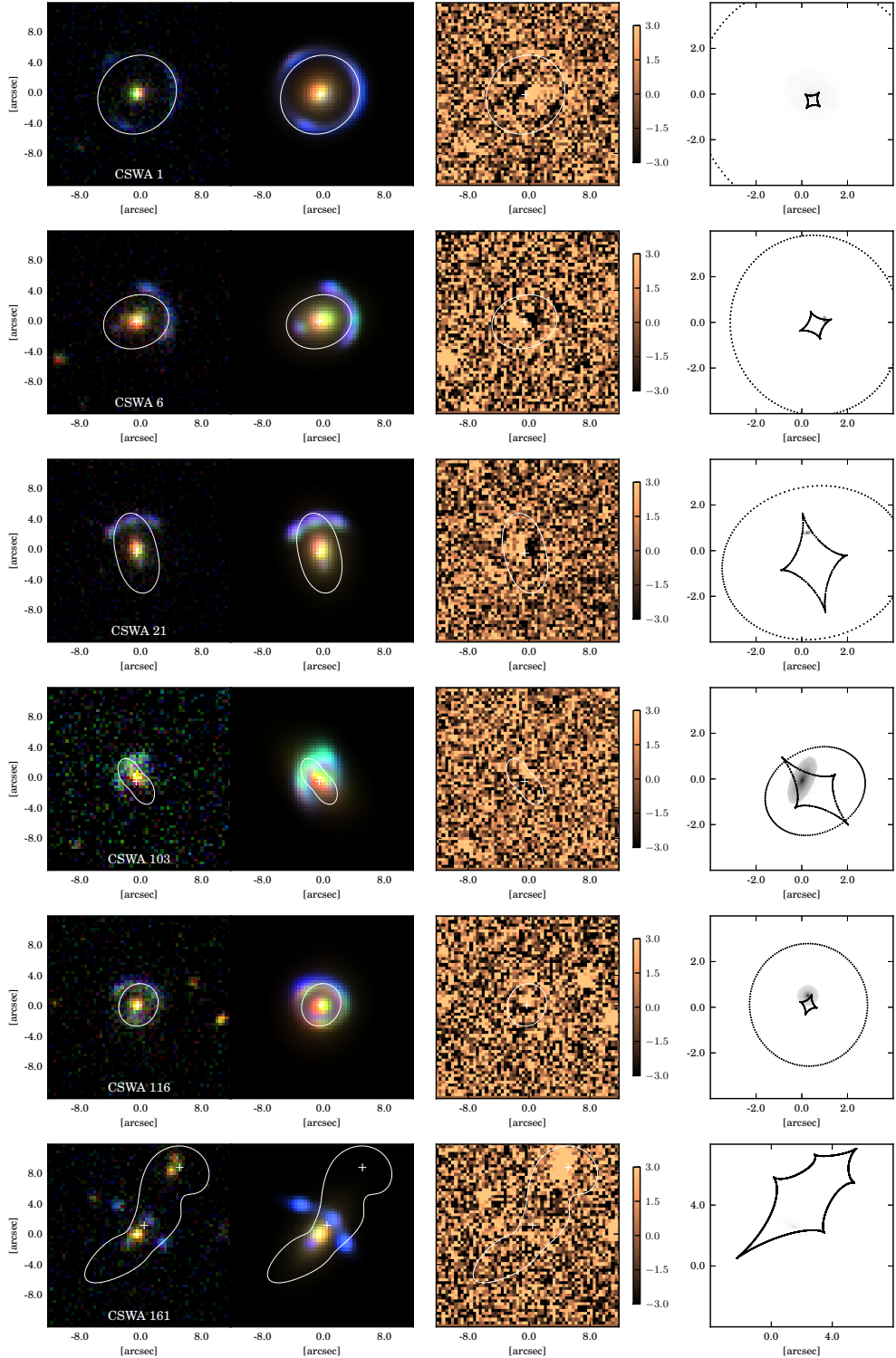}
\caption{Results of the lens modelling analysis. For each row, a system is shown, from left to right with north up and east left, composed the SDSS data in $g$, $r$, $i$ bands ($20''\times20''$), the best fit model prediction, the residuals in $g$-band ([data-model]/noise), and the associated source plane light distribution ($8''\times8''$, except CSWA~161 -- $16''\times16''$ ). The critical curves are presented in the first three panels whereas the caustics are shown in the last one. }
\label{fig:model}
\end{figure*}

\begin{figure*}
\includegraphics[trim=0mm 200mm 0mm 0mm, clip, scale=0.55]{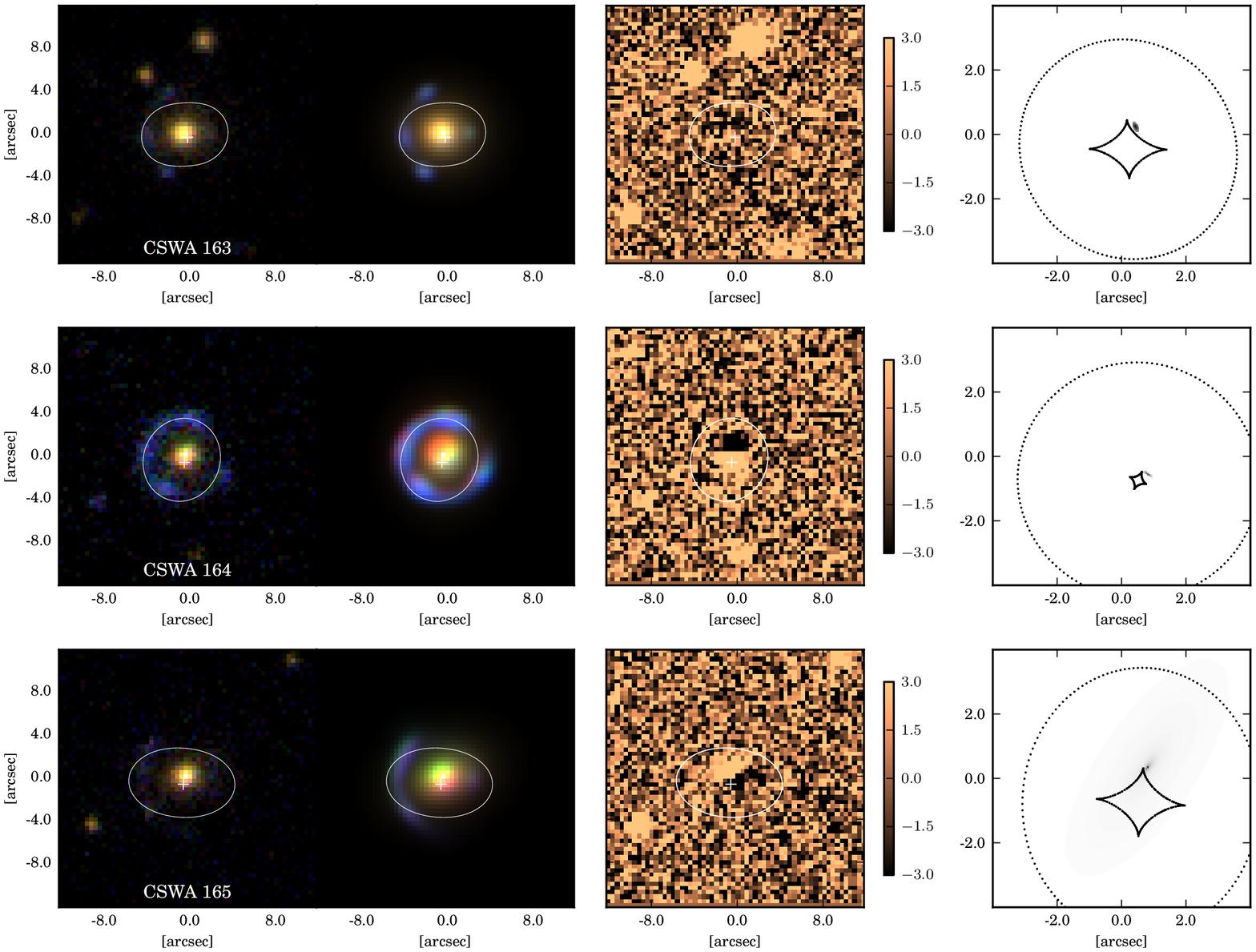}
\caption{Results of the lens modelling analysis (continued).}
\label{fig:model2}
\end{figure*}

The fitted parameters, the lens and source photometry (with 68 per cent confidence level range) and the noise parameters are listed in Tables \ref{tab:param1}, \ref{tab:param2} and \ref{tab:AAA}.
The source redshift, necessary for source properties studies, was known for eight of 9 systems from the selected sample.
The models of the 9 systems analysed in this study are shown in Figures \ref{fig:model}-\ref{fig:model2}.
We plot the original data from the SDSS, the best fitting model, the residuals and the associated source plane light distribution.
In Fig.\ \ref{fig:corr} we present the 2D projection of the likelihood in 29D-space of the fitted parameters (we only show 13 of the most important parameters) for one of the analysed systems (CSWA~21).
We use colour-coding based on $\sigma$ confidence regions defined for each pixel on 2D plane as $\sqrt{2\log(L_{max}/L)}$, where $L_{max}$ is the maximum value of the likelihood on the plane and $L$ is the value of the likelihood in a given pixel.
Diagonal panels show the likelihood distributions of each of the parameters marginalized over all other dimensions.

\subsection{Notes on individual lens models in the most interesting systems}

A couple of lens systems from our sample were previously modelled using imaging with higher resolution than SDSS. 
This gives us an opportunity to test our models obtained from lower quality data.
We can claim that the re-modelling of a few systems (CSWA 1, 6, 21) was carried out successfully.
Here we present details on these re-modelled systems for comparison with the previous models.
The difference between our and literature values of the Einstein radius in these systems is negligible, confirming the goodness of our fits.
We also describe in detail a few most interesting and well modelled systems in our new sample. 

\subsection*{ CSWA 1}
This system was discovered in the SDSS data release 5 by \citet{2007ApJ...671L...9B}.
This is an almost complete ($\sim 300^{\circ}$) Einstein ring 10 arcseconds in diameter.
The deflecting galaxy is at redshift 0.444 while the source is a star-forming galaxy at redshift 2.379.
Using the follow-up images from the 2.5 m Isaac Newton Telescope a few lens models were found with the most probable model - the power law with a volume mass density $\rho \sim r^{-1.96\pm0.02}$ and an axis ratio of $\sim0.8$ (\citealt{2008MNRAS.388..384D}).
For the SIE model with external shear the Einstein radius is $\theta_E = 5\farcs16 \pm 0\farcs06$ and the axis ratio $q_M = 0.81 \pm 0.02$

With our data quality there was no possibility to fit a reliable lens model with free power-law index.
Nevertheless, our axis ratio and ring diameter obtained from the SIE model agree with the previous study ($\theta_E = 5\farcs08 \pm 0\farcs02$ within the range of 1$\sigma$ and $q_M = 0.89^{+0.03}_{-0.04}$ within the range of 2$\sigma$).
The modelling of the lens light profile is satisfactory, comparing to previous studies.
We note that it is probably not the best approximation to describe the source with a single S\'ersic profile, hence, comparing to the previous results, our magnification value is probably somewhat underestimated.

\subsection*{ CSWA 6}
The system comprises a bright arc and one counter-image at redshift $z=2.00$ (\citealt{2009ApJ...699.1242L}).
This star-forming galaxy is strongly lensed by a foreground $z=0.422$ luminous red galaxy.
This system was found in the SDSS DR5 and was followed up with Subaru 8.2 m telescope and 3.5 m telescope at Apache Point Observatory.
A simple lens model for the system, assuming a singular isothermal ellipsoid mass distribution, yields an Einstein radius of $\theta_E = 3\farcs82 \pm 0\farcs03$.
Additional observations from Subaru telescope showed that the central lensing galaxy is accompanied by two smaller galaxies.
The magnification is estimated at $27 \pm 1$.

Our modelling recovered the known parameters of this lens only partially.
The magnification is likely to be underestimated but we obtained the Einstein radius in agreement within $1\sigma$ ($\theta_E = 3\farcs86^{+0\farcs06}_{-0\farcs04}$).
It seems that the new model is more elliptical ($q_M = 0.68^{+0.12}_{-0.14}$).
We should note that the probability histogram of parameter $q_M$ shows the possibility of the existence of two different axis ratio values with almost the same likelihood.
The lens light profile parameters are within the range of 3$\sigma$ to values found in the previous studies, probably due to the oversimplification of the fitted mass profile.
There was no possibility of modelling the parameters of neighbouring galaxies. 
The system consists of 4 lenses (\citealt{2010MNRAS.407..225V}) which we could not distinguish in our images.

\subsection*{ CSWA 21}
This lens system has been studied in detail by \citet{2007ApJ...662L..51A}. 
The system contains a Lyman Break Galaxy (LBG) at redshift $z=2.73$ (confirmed by additional spectroscopic observations) strongly lensed by a Luminous Red Galaxy ($z=0.38$).
The arc was first identified in the SDSS DR4 and later observed by 3.5 m telescope at Apache Point Observatory.

\begin{figure*}
\includegraphics{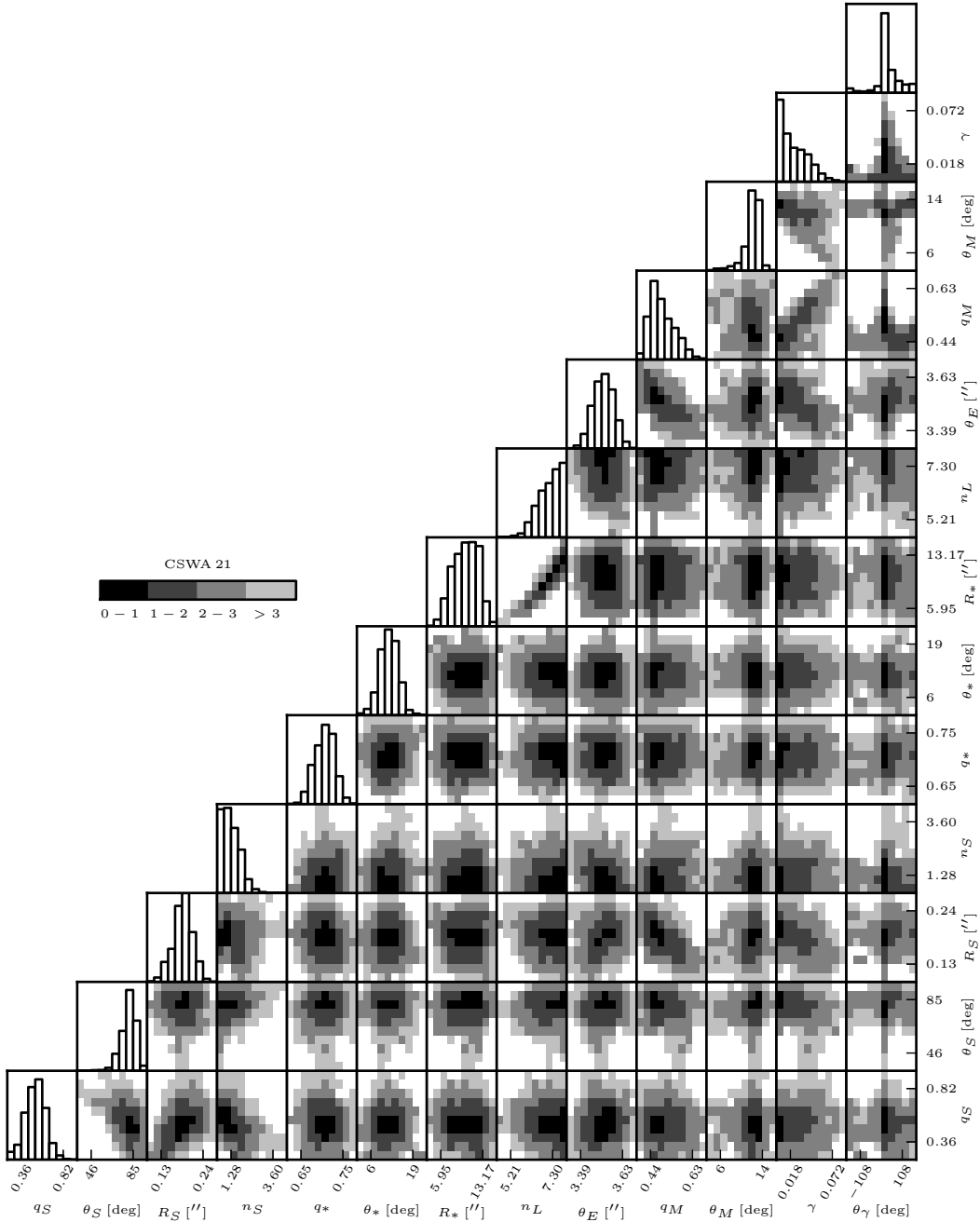}
\caption{2D projection of likelihood distributions of the 13 most important parameters of the model obtained from the MCMC sampling for CSWA 21 lensing system. Diagonal panels show likelihood distributions of each parameter marginalized over all other dimensions. We use colour-coding based on $\sigma$ confidence regions defined for each pixel on 2D plane as $\sqrt{2\log(L_{max}/L)}$, where $L_{max}$ is the maximum value of the likelihood on the plane and $L$ is the value of the likelihood in a given pixel.}
\label{fig:corr}
\end{figure*}

Using only the SDSS data we obtained results in good agreement with \citet{2007ApJ...662L..51A}.
A simple SIE + external shear model yields a good fit.
\citet{2007ApJ...662L..51A} estimated the Einstein radius of $\theta_E = 3\farcs32 \pm 0\farcs16$ and the axis ratio of $0.47 \pm 0.06$.
The magnification of the system is evaluated at $12.3^{+15}_{-3.6}$.
Our estimate for $\theta_E$ is in agreement within 2$\sigma$ and the axis ratio fits even better ($\theta_E = 3\farcs52 \pm 0\farcs06$, $q_M = 0.48^{+0.07}_{-0.04}$).
The magnification is also in a good agreement within 1$\sigma$.
The distribution of S\'ersic indices for the lens and the source, $n_L$ and $n_S$, is not symmetric (see the Fig.\ \ref{fig:corr}) and in both light models tends to the assumed limit of fitted parameter ($n_L \to 8$ and $n_S \to 0$, respectively).
The lens S\'ersic index is noticeably degenerated with the effective radius in the lens light profile.
From Fig.\ \ref{fig:corr}, we also notice a slight degeneracy between the ellipticity in the mass profile and external shear.
This is a common issue in strong lens modelling.

\subsection*{ CSWA 103}
In this system we notice an extreme value of external shear ($\gamma = 0.214^{+0.063}_{-0.081}$), although the mass profile perfectly follows the light profile.
The closer look to the system environment shows the group of objects which can be responsible for external shear.
However, we assume the contribution of those objects to the lensing is accurately taken into account through the shear.

\subsection*{ CSWA 161}
The system contains four blue images with a central lens galaxy (L1) at redshift $z = 0.497 \pm 0.029$.
The source redshift is unknown.
After a first attempt to model this system with one SIE lens we obtained a strongly flattened lens and high value of shear.
The position angles of lens and shear pointed towards another galaxy at photometric redshift near to the main lens redshift.
We therefore considered the two lens system with the second lens (L2) situated at $(\alpha,\delta) = (355\fdg11727798, 29\fdg79863137)$ at a distance of about 9 arcsec from the main lens.
The photometric redshift of the second lens, $z = 0.502 \pm 0.034$, allows us to use a single lens plane.
The mass model of the second lens galaxy is also assumed as SIE. 
In the binary lens model the axis ratio for the main lens increased, but the shear contribution decreased significantly, indicating minimal influence from the further galaxies. 
We obtain two-lens-model with close Einstein radii ($2\farcs67^{+0\farcs13}_{-0\farcs12}$ for the primary lens and $3\farcs31^{+0\farcs27}_{-0\farcs26}$ for the second lens).
For the primary lens the mass profile is not tied to the light distribution (comparing $q_M = 0.31^{+0.05}_{-0.04}$ and $q_* = 0.53^{+0.05}_{-0.05}$) what should not be unexpected in the binary lens systems.

\subsection{Lens properties}
\begin{figure}
\includegraphics{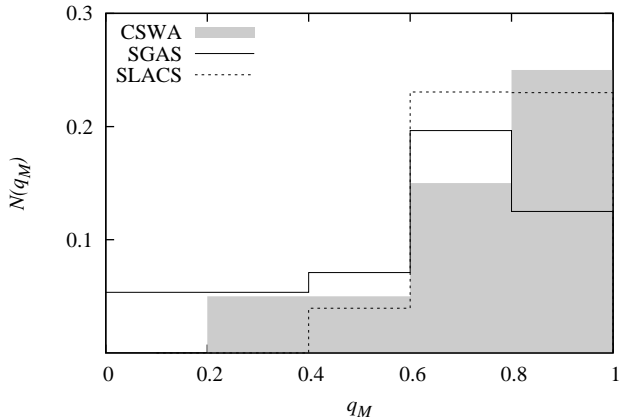}
\caption{Characteristics of the CSWA lenses. The plot presents the distribution of lens mass axis ratios for our lenses (filled histogram) and for the SGAS sample (solid line) and SLACS (dashed line). The CSWA sample shows similar axis ratios than the SGAS and SLACS lenses.}
\label{fig:hist_qSIE}
\end{figure}

Figure \ref{fig:hist_qSIE} shows the normalized histogram of SIE axis ratios with comparison to the Sloan Giant Arcs Survey (SGAS) sample presented in \citet{2012MNRAS.420.3213O} and the Sloan Lens ACS (SLACS) sample from \citet{2008ApJ...682..964B}.
Our sample shows similar axis ratios to both samples used to compare.

\begin{figure}
\includegraphics{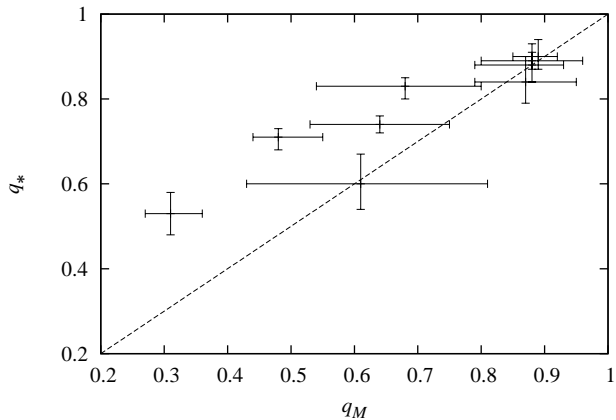}
\caption{Characteristics of the CSWA lenses. The correlation between lens axis ratio of the mass and of the light distribution. The light distribution generally follows the mass distribution in the CASSOWARY sample. For the vast majority of systems the lens mass ellipticity is slightly larger than the light ellipticity. For our CASSOWARY sample we found $\left< q_M/q_*\right> = 0.89 \pm 0.15$ which is consistent with \citet{2006ApJ...649..599K} and \citet{2012ApJ...761..170G}. }
\label{fig:qSIE_qlight}
\end{figure}

\begin{figure}
\includegraphics{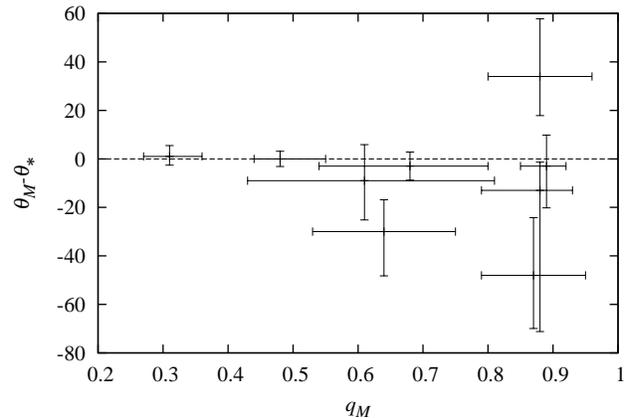}
\caption{The angular offset between the major axis of the lens mass distribution and that of the light as a function of axis ratio of the mass distribution.}
\label{fig:papa}
\end{figure}

Figure \ref{fig:qSIE_qlight} compares the ellipticity of the mass model to that of the stellar component of the lens.
For the vast majority of systems the lens mass ellipticity is slightly larger than the light ellipticity distribution.
Previous studies showed that the relation between light and mass ellipticity should be linear.
\citet{2006ApJ...649..599K} showed for the SLACS sample that $\left< q_M/q_*\right> = 0.99 \pm 0.11$, however \cite{2012ApJ...761..170G} presented for the SL2S sample significant more dispersion $\left< q_M/q_*\right> = 0.95 \pm 0.48$.
Nevertheless, for the CASSOWARY sample we found $\left< q_M/q_*\right> = 0.89 \pm 0.15$ that is not significantly different from the earlier results.
We also find that the position angles of the mass and light are well aligned (Fig. \ref{fig:papa}).

The assumption of isothermality may introduce major systematic shifts in lens parameters.
Hence, we checked the correlation between the Einstein radius, the axis ratio, the external shear and the power-law index by re-modelling the systems with power-law profiles.
We use a few fixed values of the power-law index, as the quality of the SDSS data does not enable us to model the full free power-law profile.
In Table \ref{tab:pl} in the Appendix we present the profile parameters obtained for several power-law indices. 
We find a strong degeneracy between the power-law index and the lens mass axis ratio only in profiles of systems: CSWA 6 and 21.
Hence, in these system the power-law profile should be used to obtain the most proper model. 
In other systems we observe a degeneracy of the power-law index with the value of the external shear.

In most cases we notice the degeneracy between S\'ersic index and effective radius which was expected and considered by \eg \cite{2012ApJ...761..170G}.
The S\'ersic profile is specified by two parameters: index and effective radius and we obtained these parameters strongly correlated in almost all fitted profiles (see Fig.\ \ref{fig:corr}).

\subsection{Source properties}

\begin{figure}
\includegraphics{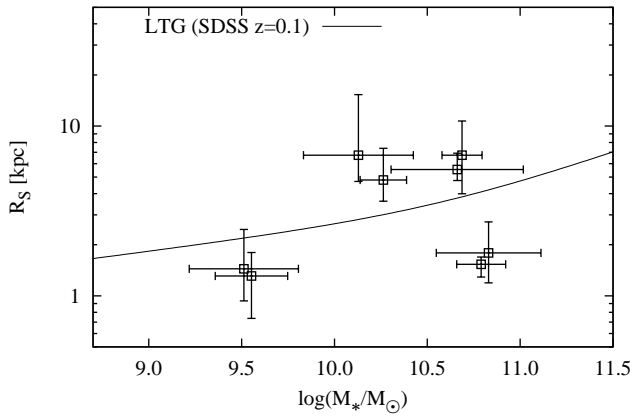}
\caption{Size-mass relation for 8 CSWA lensed source galaxies with known redshifts (large squares with error bars). For comparison we present the size-mass relation for late-type galaxies (LTG) in local Universe from \citet{2003MNRAS.343..978S} based on the SDSS galaxy sample.
The stellar masses were computed using the spectral energy distributions (SED) converted to a Chabrier IMF normalization.
The obtained masses and radii are consistent with the SDSS relation.}
\label{fig:mass_reff}
\end{figure}
\begin{figure}
\includegraphics{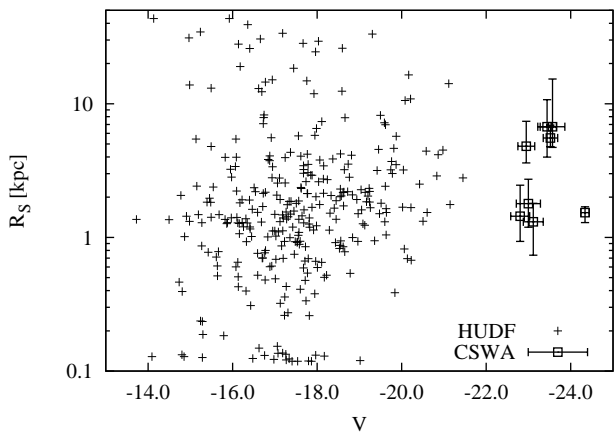}
\caption{Size-luminosity plane for 8 CSWA lensed source galaxies with known redshifts (large squares with error bars). For comparison we present the size-luminosity plane for galaxies in the HUDF catalogue (small crosses). We notice that our CSWA sample represents brighter galaxies than the HUDF sample.}
\label{fig:lum_reff}
\end{figure}

\begin{figure}
\includegraphics{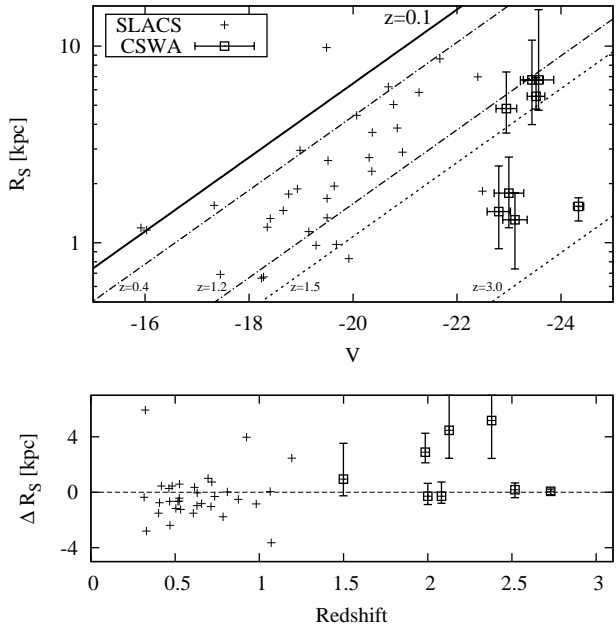}
\caption{Top panel: Size-luminosity relation for 8 CSWA lensed source galaxies with known redshifts (large squares with error bars). For comparison we present the size-luminosity relation for galaxies in the SLACS sample (small crosses). We notice that our CSWA sample represents brighter galaxies than the SLACS sample. We also plot the size-luminosity relation for different redshift (lines from left to right for redshift: 0.1, 0.4, 1.2, 1.5, and 3.0). Bottom panel: Residuals between observed effective radius of source galaxies and predicted radius from the size-luminosity evolution relation. We notice that in a few source galaxies the effective radius is greater than expected from the size-luminosity evolution.}
\label{fig:evolution}
\end{figure}

Our source sample is a set of galaxies at redshifts between 1.5 and 3.0 (Table \ref{tab:CSWAsample}).
Finally, we explore the size-stellar mass plane (Fig.\ \ref{fig:mass_reff}) and size-luminosity plane (Fig.\ \ref{fig:evolution}).
The stellar masses ($\mathrm{M_*}$) have been computed using the spectral energy distributions (SED) converted to a Chabrier IMF normalization and we find that they span a range of stellar masses typical of star-forming galaxies at these redshifts (\citealt{2012ApJ...745...85L}).
We also plot a size-mass relation for low redshift galaxies obtained by \citet{2003MNRAS.343..978S}.
It is clearly not possible to obtain a reliable parametrized stellar mass-radius relation due to the very low sample size and large uncertainties in the mass profiles.
The uncertainties in source masses are caused by using low resolution data and indirect fitting of source light profile. 
However, we can confirm that the extended sample from low resolution data will be useful in similar studies.

For comparison we use the Hubble Ultra Deep Field (HUDF) catalogue with high-redshift galaxies between z $\sim$ 1.5 and 3.0 (\citealt{2006AJ....132..926C}).
From size-luminosity plane (Fig.\ \ref{fig:lum_reff}) we can conclude that our CSWA sample complements the HUDF catalogue at redshifts 1.5-3.0 with brighter galaxies with luminosities between -22.0 and -24.5 mag (in the $V$ band). 
We investigate the evolution of the size-luminosity relation in a function of redshift based on the SLACS and CSWA source galaxies and the SDSS local galaxies.
The SLACS sample represents galaxies at redshifts range [0.4,1.2] with multiple nebular emission lines (\citealt{2006ApJ...638..703B}).
We assumed the size-luminosity relation for local galaxies from \citet{2013ApJ...777....1B}: $\log(R_S)=-0.188 V -2.95$.
For a combined sample of SLACS and CSWA source galaxies we fit this relation for different redshifts and find:
\[
\log(R_S)=-0.188 V -2.95+(z-0.1)(-0.556 \pm 0.041).
\] 
In the top panel of Figure \ref{fig:evolution} we present the size-luminosity plane with the SLACS and CSWA source samples.
We also plot the size-luminosity relation for various redshifts, including $z\sim0.1$ representing local galaxies.
In the bottom panel we plot residuals between the observed effective radius of source galaxies and the predicted radius from the size-luminosity relation.
We are not able to decide whether the pure-size evolution or pure-luminosity evolution plays a key role in the observed evolution. 
Although we can assume that the CSWA sources are the progenitors of the SLACS sources as similarly the SLACS sources are matched to the local SDSS galaxies population.
Hence, we likely observe the passive evolution of disc galaxies and the decline of star formation rate.
We caution the reader as the observational bias is probably one of the most important restriction here; we cannot exclude the influence of selection bias in the SDSS data search for lens systems.
The observed sources in the CSWA sample are brighter and larger (for example compared to the HUDF sample in the same redshift range) maybe only due to the SDSS thresholds.

\section{Conclusions}

In this paper we have presented gravitational lens models of 9 galaxy-scale lenses identified in the CASSOWARY catalogue.
For our modelling we only used the original multi-band SDSS data, hence we are able to explore the homogeneous sample of galaxy-galaxy lensing systems.
We used the S\'ersic profile to constrain the surface brightness model of the lenses and sources and the SIE profile to obtain the lens mass model.
For each system we derived the Einstein radius, axis ratio, and positional angle of the mass distribution. 
We also obtained surface brightness parameters (effective radius, axis ratio, positional angle, and S\'ersic index) for both lens and source light distributions.

The homogeneous sample allowed us to investigate the properties of the lenses and sources. 
We found that in the vast majority of our systems the lens mass follows the lens light and the alignment between the lens mass and lens light axis ratios is good. 
However, in a few cases the assumed isothermality for the mass profile may be insufficient.
In systems CSWA 6 and 21 we noticed a strong degeneracy between the power-law index and the lens axis ratio.

By exploiting the gravitational lensing effect we were able to determine sizes, stellar masses, and luminosities of a sample of faint galaxies.
The studied redshift range is crucial for understanding of galaxy evolution processes. 
Hence, we have been interested in exploring mass to radius and luminosity to radius relationships.
We obtained robust results which are encouraging for further use of low-resolution data.
However, we were not able to determine the galaxy type from S\'ersic profile indices due to large uncertainties.
Our sample probably contains mostly irregular galaxies and this is the reason that the S\'ersic indices are not reasonable in half of the studied cases.
The source could have a more complicated structure which cannot be described well by a single model.
For instance, \citet{2012ApJ...745...85L} showed that star-forming galaxy profiles at redshifts range 1.5-3.0 should be possible to model simultaneously with at least few S\'ersic profiles.
On the other hand, we notice that the CSWA sample complements the high-redshift HUDF catalogue with bright galaxies and also the CSWA sample extends the SLACS sample for higher redshift lensed galaxies.
We presented evolution of the size-luminosity relation for star-forming galaxies as a function of redshift. 
The relation was fitted using both samples - the CSWA and SLACS galaxies.
We assume that the CSWA sources are the progenitors of the SLACS sources, whereas the SLACS source are progenitors of the local SDSS galaxies.

In general, our results showed that we are still able to recover lens and source parameters even when using only the low-resolution images from the SDSS survey.
However, we caution the reader as to necessity of careful selection of modelled systems with the presence of the counter image being the most important criterion.

\section*{Acknowledgements}

We thank Jan Skowron for fruitful discussion about Figure \ref{fig:corr}.
ZKR was supported by the Polish National Science Centre under grant DEC-2011/01/N/ST9/03069.
VB acknowledges financial support from the Royal Society.

\bibliographystyle{mn2e}

\begin{appendix}
\section{Power-law}

Here we presented the results of re-modelling systems with power-law profiles.
We test the models using only a few fixed values of power-law index from range $[0.8,1.2]$.
In Table \ref{tab:pl} we show for each system from the CASSOWARY sample used power-law index and obtained: the Einstein radius $\theta_E$, the axis ratio $q_M$ and the position angle $\theta_M$ in the lens mass profile, the value of external shear $\gamma$ and the position angle of shear $\theta_{\gamma}$. 

\begin{table}
\caption{Power-law tests}
\centering
\scalebox{0.6}{
\begin{tabular}[b]{l c c c c c c c c c c c c c c c }
CSWA id & power-law index &$\theta_E~['']$ & $q_M$ &$\theta_M$ [deg] & $\gamma$ &$\theta_{\gamma}$ [deg] \\
\hline
CSWA 1 \\
&0.8 &$ 5.07 ^{+0.02 }_{-0.02 }$& $ 0.89 ^{+0.02 }_{-0.03 }$& $ 131 ^{+7 }_{-12 }$& $ 0.016 ^{+0.014 }_{-0.010 }$& $ 3 ^{+22 }_{-16 }$\\
&0.9 &$ 5.07 ^{+0.02 }_{-0.02 }$& $ 0.89 ^{+0.03 }_{-0.03 }$& $ 131 ^{+8 }_{-9 }$& $ 0.017 ^{+0.012 }_{-0.012 }$& $ -8 ^{+26 }_{-17 }$\\
&0.95 &$5.08 ^{+0.02 }_{-0.02 }$& $ 0.89 ^{+0.03 }_{-0.03 }$& $ 131 ^{+8 }_{-13 }$& $ 0.019 ^{+0.013 }_{-0.012 }$& $ -11 ^{+22 }_{-16 }$\\
&(1.0) & $ 5.08 ^{+0.02 }_{-0.02 }$& $ 0.89 ^{+0.03 }_{-0.04 }$& $ 130 ^{+8 }_{-14 }$& $ 0.021 ^{+0.012 }_{-0.012 }$& $ -14 ^{+19 }_{-13 }$\\
&1.05 &$ 5.07 ^{+0.02 }_{-0.02 }$& $ 0.89 ^{+0.04 }_{-0.04 }$& $ 127 ^{+10 }_{-15 }$& $ 0.026 ^{+0.012 }_{-0.014 }$& $ -17 ^{+15 }_{-12 }$\\
&1.1 &$ 5.07 ^{+0.03 }_{-0.02 }$& $ 0.90 ^{+0.04 }_{-0.05 }$& $ 126 ^{+11 }_{-12 }$& $ 0.028 ^{+0.011 }_{-0.013 }$& $ -20 ^{+13 }_{-12 }$\\
&1.2 &$ 5.07 ^{+0.02 }_{-0.02 }$& $ 0.91 ^{+0.05 }_{-0.07 }$& $ 125 ^{+14 }_{-16 }$& $ 0.033 ^{+0.011 }_{-0.013 }$& $ -26 ^{+14 }_{-10 }$\\
\hline
CSWA 6\\
&0.8 &$ 3.83 ^{+0.04 }_{-0.02 }$& $ 0.84 ^{+0.04 }_{-0.04 }$& $ 102 ^{+4 }_{-7 }$& $ 0.022 ^{+0.012 }_{-0.012 }$& $ 135 ^{+18 }_{-12 }$\\
&0.9 &$ 3.84 ^{+0.04 }_{-0.03 }$& $ 0.78 ^{+0.06 }_{-0.06 }$& $ 103 ^{+4 }_{-6 }$& $ 0.024 ^{+0.016 }_{-0.015 }$& $ 140 ^{+23 }_{-13 }$\\
&0.95 &$ 3.85 ^{+0.04 }_{-0.04 }$& $ 0.75 ^{+0.08 }_{-0.11 }$& $ 103 ^{+4 }_{-4 }$& $ 0.029 ^{+0.016 }_{-0.015 }$& $ 138 ^{+31 }_{-16 }$\\
&(1.0) & $ 3.86 ^{+0.06 }_{-0.04 }$& $ 0.68 ^{+0.12 }_{-0.14 }$& $ 103 ^{+3 }_{-3 }$& $ 0.037 ^{+0.017 }_{-0.019 }$& $ 152 ^{+21 }_{-26 }$\\
&1.05 &$ 3.89 ^{+0.05 }_{-0.05 }$& $ 0.54 ^{+0.16 }_{-0.06 }$& $ 103 ^{+2 }_{-3 }$& $ 0.045 ^{+0.019 }_{-0.018 }$& $ 166 ^{+10 }_{-29 }$\\
&1.1 &$ 3.90 ^{+0.06 }_{-0.05 }$& $ 0.49 ^{+0.12 }_{-0.09 }$& $ 103 ^{+2 }_{-2 }$& $ 0.052 ^{+0.028 }_{-0.023 }$& $ 171 ^{+7 }_{-26 }$\\
&1.2 &$ 4.00 ^{+0.08 }_{-0.09 }$& $ 0.31 ^{+0.08 }_{-0.05 }$& $ 104 ^{+1 }_{-2 }$& $ 0.082 ^{+0.025 }_{-0.027 }$& $ 175 ^{+4 }_{-8 }$\\
\hline
CSWA 21\\
&0.8 &$ 3.56 ^{+0.05 }_{-0.05 }$& $ 0.59 ^{+0.02 }_{-0.03 }$& $ 12 ^{+1 }_{-2 }$& $ 0.007 ^{+0.016 }_{-0.006 }$& $ 73 ^{+42 }_{-60 }$\\
&0.9 &$ 3.54 ^{+0.05 }_{-0.05 }$& $ 0.53 ^{+0.03 }_{-0.02 }$& $ 13 ^{+1 }_{-1 }$& $ 0.008 ^{+0.012 }_{-0.006 }$& $ 39 ^{+75 }_{-46 }$\\
&0.95 &$ 3.52 ^{+0.06 }_{-0.06 }$& $ 0.51 ^{+0.07 }_{-0.04 }$& $ 12 ^{+1 }_{-3 }$& $ 0.016 ^{+0.033 }_{-0.012 }$& $ 27 ^{+61 }_{-20 }$\\
&(1.0) & $ 3.52 ^{+0.06 }_{-0.06 }$& $ 0.48 ^{+0.07 }_{-0.04 }$& $ 12 ^{+1 }_{-1 }$& $ 0.017 ^{+0.024 }_{-0.013 }$& $ 23 ^{+64 }_{-15 }$\\
&1.05 &$ 3.46 ^{+0.10 }_{-0.06 }$& $ 0.54 ^{+0.09 }_{-0.12 }$& $ 12 ^{+1 }_{-1 }$& $ 0.057 ^{+0.027 }_{-0.047 }$& $ 15 ^{+39 }_{-6 }$\\
&1.1 &$ 3.38 ^{+0.13 }_{-0.05 }$& $ 0.68 ^{+0.10 }_{-0.23 }$& $ 12 ^{+2 }_{-4 }$& $ 0.107 ^{+0.024 }_{-0.063 }$& $ 15 ^{+11 }_{-3 }$\\
&1.2 &$ 3.34 ^{+0.06 }_{-0.05 }$& $ 0.76 ^{+0.10 }_{-0.16 }$& $ 11 ^{+3 }_{-5 }$& $ 0.149 ^{+0.021 }_{-0.034 }$& $ 14 ^{+3 }_{-2 }$\\
\hline
CSWA103\\
&0.9 & $ 1.91 ^{+0.17 }_{-0.13 }$& $ 0.57 ^{+0.20 }_{-0.17 }$& $ 37 ^{+12 }_{-13 }$& $ 0.166 ^{+0.069 }_{-0.090 }$& $ 54 ^{+18 }_{-9 }$\\
&(1.0) & $ 1.87 ^{+0.14 }_{-0.12 }$& $ 0.61 ^{+0.20 }_{-0.18 }$& $ 34 ^{+14 }_{-15 }$& $ 0.214 ^{+0.063 }_{-0.081 }$& $ 51 ^{+9 }_{-7 }$\\
&1.1 & $ 1.84 ^{+0.13 }_{-0.12 }$& $ 0.68 ^{+0.20 }_{-0.21 }$& $ 32 ^{+18 }_{-17 }$& $ 0.262 ^{+0.057 }_{-0.077 }$& $ 50 ^{+7 }_{-5 }$\\
\hline
CSWA 116 \\
&0.9 & $ 2.62 ^{+0.03 }_{-0.03 }$& $ 0.90 ^{+0.07 }_{-0.07 }$& $ 103 ^{+37 }_{-17 }$& $ 0.094 ^{+0.030 }_{-0.021 }$& $ 169 ^{+6 }_{-7 }$\\
&(1.0) & $ 2.62 ^{+0.03 }_{-0.03 }$& $ 0.87 ^{+0.08 }_{-0.08 }$& $ 99 ^{+22 }_{-19 }$& $ 0.110 ^{+0.031 }_{-0.024 }$& $ 169 ^{+6 }_{-6 }$\\
&1.1 & $ 2.62 ^{+0.03 }_{-0.03 }$& $ 0.88 ^{+0.09 }_{-0.08 }$& $ 88 ^{+19 }_{-23 }$& $ 0.114 ^{+0.024 }_{-0.020 }$& $ 166 ^{+5 }_{-6 }$\\
\hline
CSWA 161\\
L1&0.9 & $ 2.79 ^{+0.16 }_{-0.13 }$& $ 0.41 ^{+0.05 }_{-0.04 }$& $ 142 ^{+2 }_{-2 }$& $ 0.045 ^{+0.021 }_{-0.022 }$& $ -5 ^{+13 }_{-9 }$\\
L2&(1.0) & $ 2.93 ^{+0.44 }_{-0.47 }$& $ 0.92 ^{+0.06 }_{-0.11 }$& $ 153 ^{+22 }_{-26 }$&\\
L1&(1.0) & $ 2.67 ^{+0.13 }_{-0.12 }$& $ 0.31 ^{+0.05 }_{-0.04 }$& $ 144 ^{+2 }_{-2 }$& $ 0.027 ^{+0.020 }_{-0.017 }$ &$ -5 ^{+7 }_{-7 }$\\
L2&(1.0) & $ 3.31 ^{+0.27 }_{-0.26 }$& $ 0.81 ^{+0.11 }_{-0.12 }$ &$ 133 ^{+13 }_{-14 }$ \\
L1&1.1 & $ 2.63 ^{+0.16 }_{-0.11 }$& $ 0.37 ^{+0.06 }_{-0.07 }$& $ 142 ^{+4 }_{-3 }$& $ 0.092 ^{+0.045 }_{-0.042 }$& $ -18 ^{+9 }_{-6 }$\\
L2&(1.0) & $ 3.69 ^{+0.48 }_{-0.41 }$& $ 0.89 ^{+0.08 }_{-0.12 }$& $ 135 ^{+17 }_{-27 }$& \\
\hline
CSWA 163\\
&0.9 & $ 3.40 ^{+0.05 }_{-0.05 }$& $ 0.80 ^{+0.10 }_{-0.09 }$& $ 100 ^{+10 }_{-8 }$& $ 0.097 ^{+0.022 }_{-0.027 }$& $ -97 ^{+6 }_{-6 }$\\
&(1.0) & $ 3.36 ^{+0.04 }_{-0.05 }$& $ 0.88 ^{+0.08 }_{-0.08 }$& $ 125 ^{+23 }_{-15 }$& $ 0.146 ^{+0.018 }_{-0.015 }$& $ -96 ^{+5 }_{-5 }$\\
&1.1 & $ 3.37 ^{+0.04 }_{-0.04 }$& $ 0.85 ^{+0.10 }_{-0.09 }$& $ 108 ^{+18 }_{-14 }$& $ 0.148 ^{+0.017 }_{-0.022 }$& $ -95 ^{+4 }_{-5 }$\\
\hline
CSWA 164\\
&0.9 & $ 3.68 ^{+0.02 }_{-0.02 }$& $ 0.89 ^{+0.03 }_{-0.05 }$& $ 165 ^{+7 }_{-6 }$& $ 0.017 ^{+0.010 }_{-0.010 }$& $ -50 ^{+15 }_{-20 }$\\
&(1.0) & $ 3.68 ^{+0.02 }_{-0.02 }$& $ 0.88 ^{+0.05 }_{-0.09 }$& $ 164 ^{+6 }_{-5 }$& $ 0.019 ^{+0.010 }_{-0.008 }$& $ -44 ^{+13 }_{-29 }$\\
&1.1 & $ 3.68 ^{+0.02 }_{-0.02 }$& $ 0.90 ^{+0.07 }_{-0.09 }$& $ 175 ^{+11 }_{-12 }$& $ 0.031 ^{+0.010 }_{-0.010 }$& $ -43 ^{+14 }_{-16 }$\\
\hline
CSWA 165\\
&0.9 & $ 3.70 ^{+0.18 }_{-0.19 }$& $ 0.68 ^{+0.14 }_{-0.19 }$& $ 74 ^{+18 }_{-18 }$& $ 0.132 ^{+0.056 }_{-0.084 }$& $ -67 ^{+27 }_{-11 }$\\
&(1.0) & $ 3.77 ^{+0.11 }_{-0.16 }$& $ 0.64 ^{+0.11 }_{-0.11 }$& $ 83 ^{+13 }_{-18 }$& $ 0.092 ^{+0.068 }_{-0.065 }$& $ -63 ^{+18 }_{-11 }$\\
&1.1 & $ 3.81 ^{+0.11 }_{-0.13 }$& $ 0.59 ^{+0.13 }_{-0.14 }$& $ 88 ^{+12 }_{-17 }$& $ 0.069 ^{+0.076 }_{-0.051 }$& $ -66 ^{+41 }_{-17 }$\\
\hline
\end{tabular}
}
\label{tab:pl}
\end{table}

\end{appendix}

\label{lastpage}

\end{document}